# A method for volume stabilization of single, dye-doped water microdroplets with femtoliter resolution

A. Kiraz, A. Kurt, and M. A. Dündar

*Department of Physics, Koç University, Rumelifeneri Yolu, 34450 Sariyer, Istanbul, Turkey*

M. Y. Yüce , and A. L. Demirel

*Department of Chemistry, Koç University, Rumelifeneri Yolu, 34450 Sariyer, Istanbul, Turkey*

## Abstract

A self-control mechanism that stabilizes the size of Rhodamine B-doped water microdroplets standing on a superhydrophobic surface is demonstrated. The mechanism relies on the interplay between the condensation rate that was kept constant and evaporation rate induced by laser excitation which critically depends on the size of the microdroplets. The radii of individual water microdroplets (>5 µm) stayed within a few nanometers during long time periods (up to 455 seconds). By blocking the laser excitation for 500 msec, the stable volume of individual microdroplets was shown to change stepwise.

*OCIS codes:* 010.7340, 300.2530, 010.1110

## 1. INTRODUCTION

With their almost spherical geometries and smooth surfaces, liquid microdroplets are naturally attractive to function as optical microcavities. They host high quality whispering gallery modes (WGMs) which inspired various applications in areas as: laser diagnostics[1, 2], atmospheric science, biology, and interfacial chemistry. Size control has been an important challenge in applications using microdroplets of liquids with relatively high vapor pressures. This hindered detailed studies of the dynamics at specific gas-liquid interfaces. Only very recently single water microdroplets were analyzed for long time periods[3, 4]. In these experiments, microdroplets with initial salt concentrations between 0.04-1.28 M reached to a stable volume by evaporation in a humidity controlled chamber.

We have previously demonstrated the tunability of WGMs of water microdroplets on superhydrophobic surfaces in a large spectral window[5]. Here, we show that the observed volume stability of dye-doped water microdroplets is due to the ambient humidity and size dependent laser absorption, and we introduce a mechanism that allows the volume of microdroplets to be changed stepwise without the need for any complex position control scheme such as optical tweezing[3, 4] or electric field[6] trapping.

## 2. DESCRIPTION OF THE VOLUME STABILIZATION MECHANISM

Scattering of a plane electromagnetic wave by a dielectric sphere is explained by Lorenz - Mie theory. Derivation of Lorenz - Mie theory has been extended to incorporate an incident focused Gaussian beam[7, 8]. Enhanced internal field intensities have been predicted from these analyses when resonance conditions (spectral, spatial, and polarization) are met between the excitation laser beam and specific WGMs. The resulting enhanced absorption was demonstrated using Rhodamine 6G doped ethanol microdroplets in air[9]. Microdroplets having a critical size showed much larger absorption efficiencies compared to microdroplets with diameters differing by 4 nm.

Size dependent absorption phenomenon plays a crucial role in the realization of the reported high precision volume stabilization mechanism. Figure 1 shows the calculated absorption efficiency ($Q_{abs}$) and modified absorption efficiency ($\tilde{Q}_{abs}$) as a function of radius ($R$) for a sphere having a refractive index equal to that of 50 µM Rhodamine B-doped water ($n=1.33+4\cdot10^{-5}i$)[10]. In Fig. 1a, calculation results are presented for a plane wave excitation ($\lambda$ = 532 nm, transverse polarized), where TE and TM modes appear. Fig. 1b-c simulate our experiment, with a tightly focused Gaussian beam ($\lambda$ = 532 nm, $\omega_0$ = 250 nm, linearly polarized along the x direction), and were calculated using the localized model developed by Gouesbet et al.[11] with an algorithm introduced by Lock[12]. $Q_{abs}$ and $\tilde{Q}_{abs}$ are the ratios of the total power absorbed by the sphere to the power incident upon the projected area of the sphere and to the total power of the incident beam respectively. In Fig. 1a the first order modes are suppressed due to the high absorption coefficient. For an absorbing sphere, the suppression of the higher quality lower order modes with increasing absorbance of the dielectric medium has also been reported previously[13-15]. In Fig. 1b-c the focal points are located at 6200 nm and 5800 nm (off axis illumination) respectively. In these figures only the TE modes are excited as a result of the polarization direction. As the excitation spot approaches to the surface of the sphere, absorption of the lower-quality third order modes dominate over the absorption of the second order modes.

The size of the microdroplet is determined by the balance between the condensation rate $\Gamma_{cond}$, and the evaporation rate $\Gamma_{evap}$. In our experiments, $\Gamma_{cond}$ was kept at a constant value as indicated by the dashed line in Fig. 1c. Because the self-control stability mechanism was dominant near a resonance, the factors affecting evaporation kinetics[16] other than the laser induced heating stayed nearly constant, achieving steady state conditions. Furthermore, within the range of the experimental parameters, we have observed an almost linear relationship between the laser induced heating and $\Gamma_{evap}$. Thus, in our experiments $\Gamma_{evap}$ was almost proportional to $\tilde{Q}_{abs}$ ($\Gamma_{evap} \propto I\tilde{Q}_{abs}$, $I$ is the laser intensity) and changed with the size of microdroplet as it shrunk ($\Gamma_{cond} < \Gamma_{evap}$) or grew ($\Gamma_{cond} > \Gamma_{evap}$) in size. For each WGM, the microdroplet can be at equilibrium at two different sizes as indicated by a solid square on the left of the peak and a solid circle on the right. The square corresponds to a stable equilibrium point where a self-locking mechanism is in effect: an increase in size due to condensation or a decrease in size due to evaporation is counterbalanced by a corresponding increase or decrease, respectively, in $\tilde{Q}_{abs}$ maintaining the equality. On the contrary, the circle corresponds to an unstable equilibrium: an increase in size causes $\tilde{Q}_{abs}$ to decrease and the microdroplet continues growing in size. Similarly, a decrease in size causes $\tilde{Q}_{abs}$ to increase and the microdroplet continues shrinking in size.

## 3. SURFACE PREPARATION AND EXPERIMENTAL SETUP

An ultrasonic nebulizer sprayed 50 μM Rhodamine B-doped water microdroplets (diameters between ~ 1-30 μm) into a home-built current controlled mini humidity chamber[5]. Generated microdroplets landed on a superhydrophobic surface where they rested with a nearly spherical shape. A nichrome wire resistor attached to the liquid reservoir of the nebulizer allowed for control of the ambient humidity, and hence condensation rate of water microdroplets. Superhydrophobic surfaces were prepared by spin coating hydrophobically coated silica nanoparticles (Degussa AG, Aeroxide LE1) on thin glass substrates from 50 mg/ml dispersions in ethanol[17]. Resulting films had nanometer scale surface roughness and were transparent to visible light. The average contact angle of 2 mm diameter water droplets on these surfaces was measured to be 152.6 °.

Single microdroplets were excited with a continuous wave green laser (λ = 532 nm, s-polarized) within resolution limited spots located away from the center in the direction shown as x in Fig. 1 and in the inset of Fig. 2. For microdroplets discussed in Fig. 2-4, the laser focus was positioned in the vicinity of the microdroplet's rim while for the microdroplet discussed in Fig. 5 the laser focus was positioned at almost half-radius. Upon excitation, Rhodamine B in the microdroplets served for two purposes: (i) as the absorbing agent for volume stabilization, and (ii) as the fluorescent probe enabling the observation of the WGMs in the fluorescent spectra collected from individual water microdroplets[18]. Changes in the microdroplet volume were monitored through these WGMs. A high numerical aperture microscope objective (60x, NA=1.4) was used in the inverted geometry both for excitation and collection of the fluorescence. The collected fluorescence was transmitted through a dichroic mirror, a 1.5x magnifier element, and was dispersed at a 50 cm monochromator (spectral resolution of 0.07 nm), then measured by a CCD camera. In all the contour plots presented in this report, the total time elapsed during the acquisition of each fluorescence spectrum was 3.14 sec (1 sec exposure time followed by 2.14 sec readout time).

## 4. RESULTS

### A. Volume Stabilization

Fig. 2 shows the contour plot of consecutive fluorescent spectra taken from an individual microdroplet (radius of 5.7 μm before the first acquisition) exhibiting volume stabilization. In all contour plots in this report, the intensity variation is given and background Rhodamine B emission that is not resonant with any WGMs is taken to be zero. At acquisition 16, $\Gamma_{evap}$ equals to $\Gamma_{cond}$ at a stable equilibrium point, initiating the stabilization mechanism. Starting at this point, despite any change in the ambient humidity, the WGMs stop drifting and remain stable within a full width at half maximum (FWHM) of 0.13 nm until the end of the experimental period. The observed spectral linewidth of 0.13 nm corresponds to volume stabilization within 0.53 fl (radius stabilization of 1.3 nm) over the course of the experiment (455 seconds). In this experiment, the microdroplet was excited in the very vicinity of its rim as shown in the inset of Fig. 2. However, experimental setup was unable to point the exact location of the laser focus, which demands higher spatial resolution. Simulations in Fig. 1b-c are assumed to be good representations to rationalize our observations.

**B. Stepwise Change in the Stable Volume**

By blocking the laser for short time intervals, the stable volume of a microdroplet can be changed stepwise. This is demonstrated in Fig. 3 where laser beam is blocked for 500 ms before acquisitions 10, 20, 30, 50, and 68. Each interim laser blocking causes the WGMs to start drifting to larger wavelengths. The microdroplet volume then stabilizes at consecutive stable equilibrium points. Accurate identifications of the polarization, mode numbers (n), and orders (r) of the WGMs are omitted in Fig. 3 and others to avoid errors caused by nonspherical microdroplet geometry. Since the spherical deformation of the microdroplet is mainly in the direction perpendicular to the substrate, we believe that the high quality WGMs circulate parallel to the substrate, in the equatorial plane. In this plane the cross-section of the microdroplet was still nearly circular. Hence the WGMs rotating in the clockwise and counterclockwise directions were still nearly degenerate within the resolution limit of the spectrometer. For all microdroplets discussed in this article the equatorial radius of the microdroplet is measured from the fluorescence images. Radius of the microdroplet discussed in Fig. 3 is determined to be 5.4 μm before the first acquisition.

In Fig. 3 an almost constant spectral drift of 11 nm is observed between consecutive stable equilibrium points. Assuming an ideal spherical geometry, this corresponds to an increase in the radius of this microdroplet by 102 nm, implying a volume step of 37 fl.

Because of their relatively high quality factors, here we consider that the most intense WGMs observed in the emission spectra have mode orders of 1 (TE or TM polarized). We note that even though the relatively high dye absorption suppresses the first order WGMs in the absorption efficiency spectrum (shown in Fig. 1), the emission spectrum is largely unaffected by the much smaller dye absorption at the emission wavelengths. These spectral observations do not allow for an accurate assignment of the WGMs absorbing at $\lambda$ = 532 nm. Should the absorbing WGMs be of mode orders 2 or 3, spectral steps of 7.3 nm (7.4 nm) or 7.5 nm (7.5 nm) would be observed in the first order TE (TM) WGMs in the emission spectra. These values are still largely different than the observed 11 nm spectral step, mainly due to the characteristic geometry of the microdroplets on a superhydrophobic surface[5].

## C. Other Operation Regimes

In addition to volume stabilization, two operation regimes can be observed. If $\Gamma_{cond}$ is smaller than $\Gamma_{evap}$ for any given radius, evaporation is observed in the microdroplet. For this case, the microdroplet will evaporate at an almost constant rate until the laser reaches the spectral vicinity of a WGM. When the laser resonates with a WGM, an increase in local heating will follow the enhanced laser absorption. This will result with an increase in the rate of microdroplet shrinking as observed in acquisitions 18 and 19 in Fig. 4.

When $\Gamma_{cond}$ is larger than $\Gamma_{evap}$ for all radii of a given microdroplet, it will keep growing in size as shown in Fig. 5a (WGMs drift at a rate of 0.93 nm/acquisition). Both the relatively large $\Gamma_{cond}$, and the position of the focal spot relative to the microdroplet play roles in preventing the onset of volume stabilization in Fig. 5a. Volume stabilization is observed when the laser intensity is increased from 2.5 µW to 3.0 µW as seen between acquisitions 1-5 in Fig. 5b. After the interim laser blocking for 500 ms before acquisition 6, WGMs start drifting at a rate of 0.63 nm/acquisition between acquisitions 6-13. The rate at which the size of this microdroplet grows is largely reduced between acquisitions 14-15, resulting in no spectral drift in WGMs within the spectral resolution of the experimental setup. However, volume stabilization is not established at the end of acquisition 15. WGMs drift at a rate of 0.65 nm/acquisition between acquisitions 16-30.

## 5. DISCUSSION

Selections of a proper liquid, dye concentration, and laser intensity are decisive in achieving volume stabilization using the scheme described. Liquid's physical properties come into play in relating the laser induced heating to the evaporation rate of the microdroplet. Such a detailed analysis requires the solution of the heat transfer over the three-dimensional geometry of the microdroplet. Our experiments revealed that microdroplets of water doped with 20 µM (reported in Ref. [5]) - 50 µM (reported here) Rhodamine B reflect the changes in the local heating to the microdroplet growing rate with high sensitivity. This is demonstrated by the strong laser intensity dependence of the drift rate of WGMs in Fig. 5a-b. In the absence of an absorption resonance, the spectral drift rate of WGMs is measured to decrease from 0.93 nm/acquisition to 0.63 nm/acquisition when laser intensity is increased from 2.5 µW to 3.0 µW. Hence, at these laser intensities, an increase in the local heating by less than a factor of two can be enough to achieve volume stabilization. Absorption resonances can readily provide such variations in local heating when the microdroplet is excited near its rim.

Though we have routinely achieved volume stabilization, the maximum time during which stabilization was observed did not exceed 455 seconds in our experiments. This is mainly due to photobleaching of Rhodamine B molecules. Under our experimental conditions, emission intensity quickly dropped as it is evident in Fig. 2. The corresponding change in the absorption coefficient often resulted in the loss of volume stabilization, typically after several minutes. We envision that using more photostable dye molecules or chemically synthesized quantum dots, the maximum volume stabilization time could be drastically improved.

**ACKNOWLEDGEMENT**

This work was supported by the Scientific and Technological Research Council of Turkey (Grant No. TÜBİTAK-105T500). The authors thank F. Menzel for providing the silica nanoparticles, and the Alexander von Humboldt Foundation for equipment donation. A. L. Demirel and A. Kiraz acknowledge the financial support of the Turkish Academy of Sciences in the framework of the Young Scientist Award program (Grants No. EA/TÜBA-GEBİP/2001-1-1 and A.K/TÜBA-GEBİP/2006-19).

# Figure Captions

Figure 1: Calculated absorption efficiency ($Q_{abs}$) and modified absorption efficiency ($\tilde{Q}_{abs}$) as a function of radius ($R$) for a sphere having refractive index, $n=1.33+4 \cdot 10^{-5}i$. (a) Plane wave illumination with transverse polarization. (b,c) Focused Gaussian beam illumination with linear polarization along x direction. Focus is positioned away from the center of the sphere along x direction at a distance of 6200 nm (b) and 5800 nm (c). All the waves propagate along z direction.

Figure 2: (Color online) Contour plots show the consecutive emission spectra taken from a microdroplet demonstrating volume stabilization at acquisition 16. Radius of the microdroplet is 5.7 µm before acquisition 1. Excitation intensity is 2.7 µW. Focus is positioned in the vicinity of the rim away from the center along x direction. Intensity values in arbitrary units increase from blue to red. Each consecutive segment of 20 acquisitions are separated by <1 sec period. The line plot shows the fluorescence spectrum obtained by averaging the spectra between acquisitions 16 to 160. Inset shows the fluorescence image of the microdroplet.

Figure 3: (Color online) Contour plots of emission spectra taken from a microdroplet demonstrating volume stabilization at consecutive stable equilibrium points. Radius of the microdroplet is 5.4 µm before acquisition 1. Excitation intensity is 1.5 µW. Focus is positioned in the vicinity of the rim away from the center along x direction. Intensity values in arbitrary units increase from blue to red. Each consecutive segment of 40 acquisitions are separated by <1 sec period. Arrow indicates the direction of spectral drift between consecutive stable equilibrium points.

Figure 4: (Color online) Contour plot of emission spectra from a continuously shrinking microdroplet. Radius of the microdroplet is 8.9 µm before acquisition 1. Excitation intensity is 1.9 µW. Focus is positioned in the vicinity of the rim away from the center along x direction. Intensity values in arbitrary units increase from blue to red.

Figure 5: (Color online) Contour plots of emission spectra from a microdroplet growing continuously in size. Radius is determined to be 5.2 µm before acquisition 1 in (a). Focus is positioned away from the center along x direction at a distance of 2.6 µm. Excitation intensities are 2.5 µW and 3.0 µW in (a) and (b) respectively. Intensity values in arbitrary units increase from blue to red.

Figure 1:

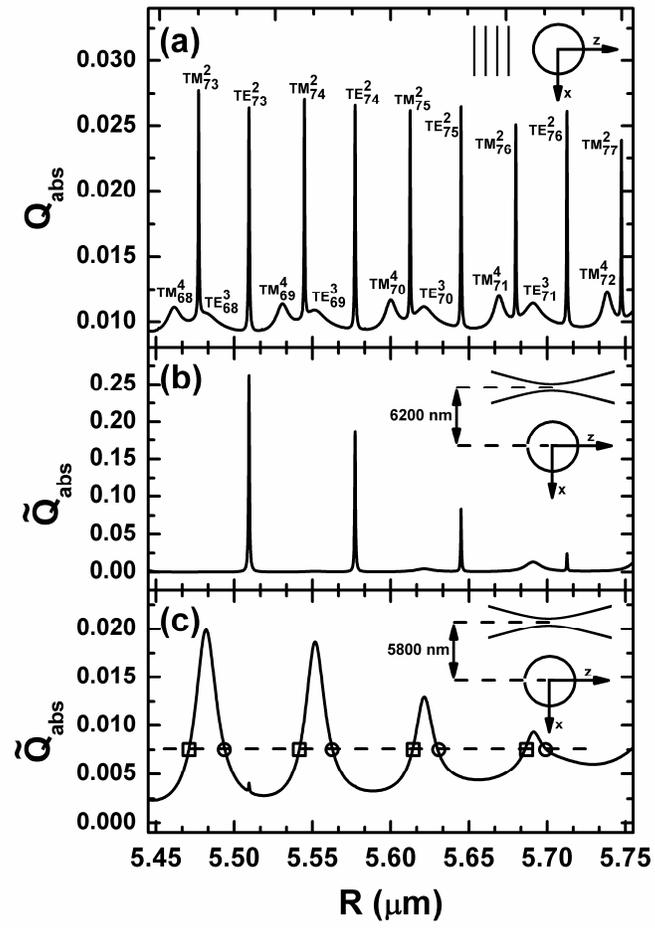

Figure 2:

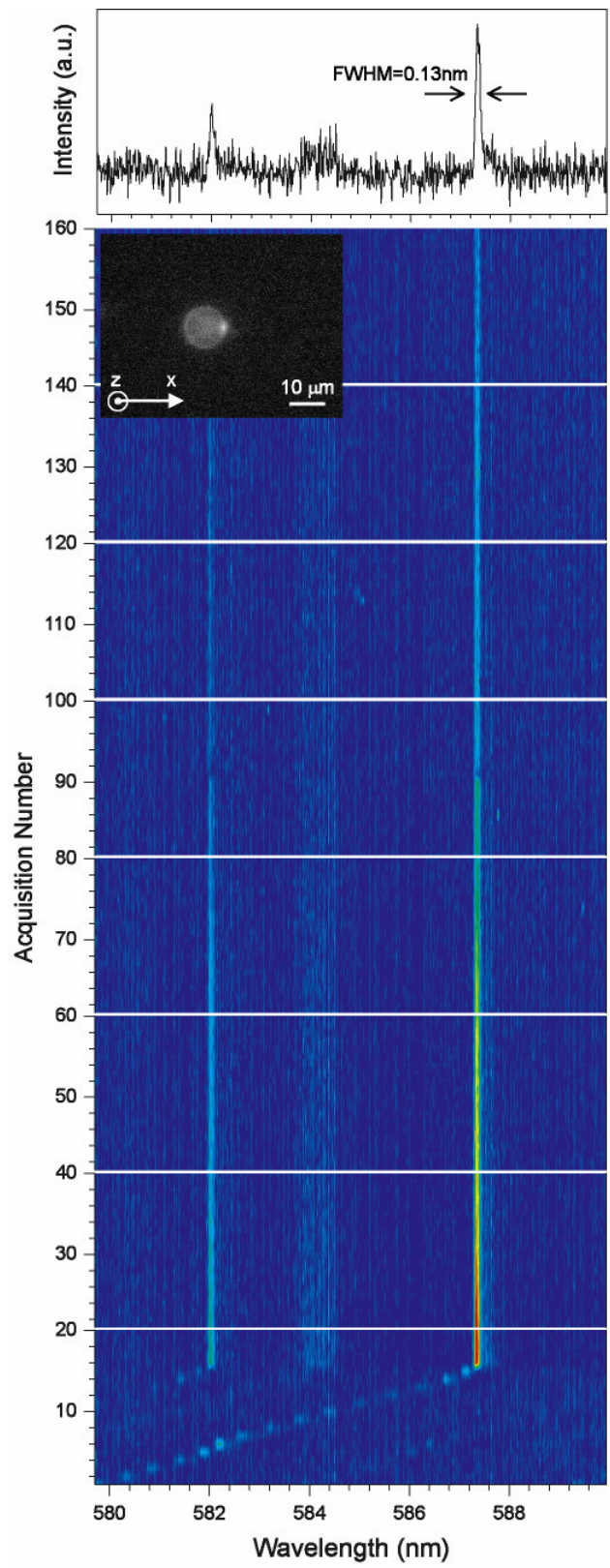

Figure 3:

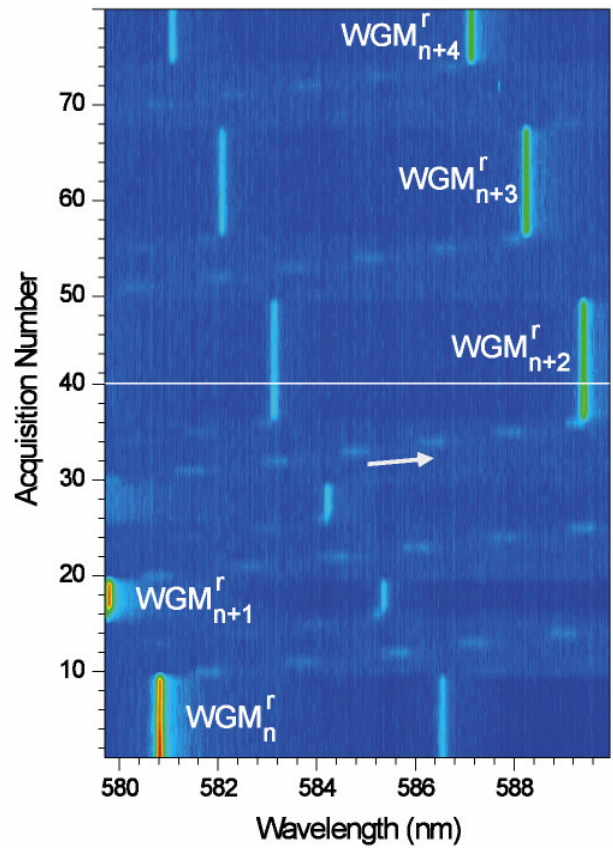

Figure 4:

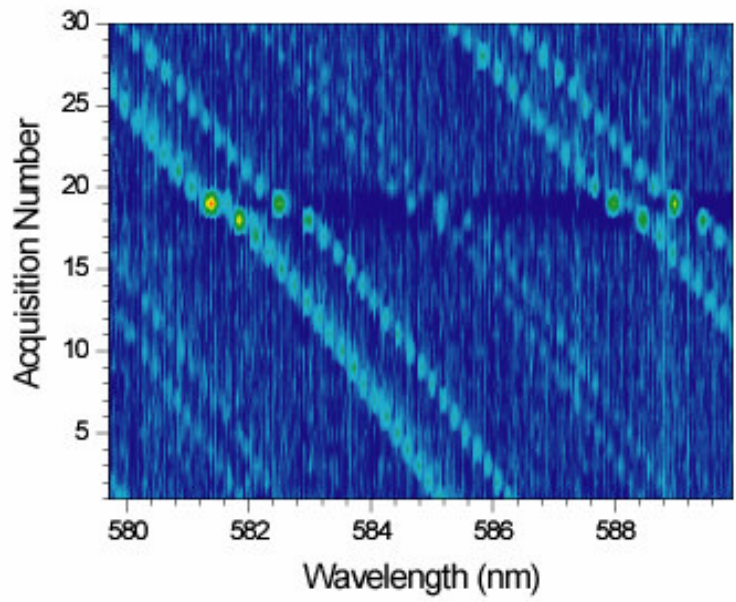

Figure 5:

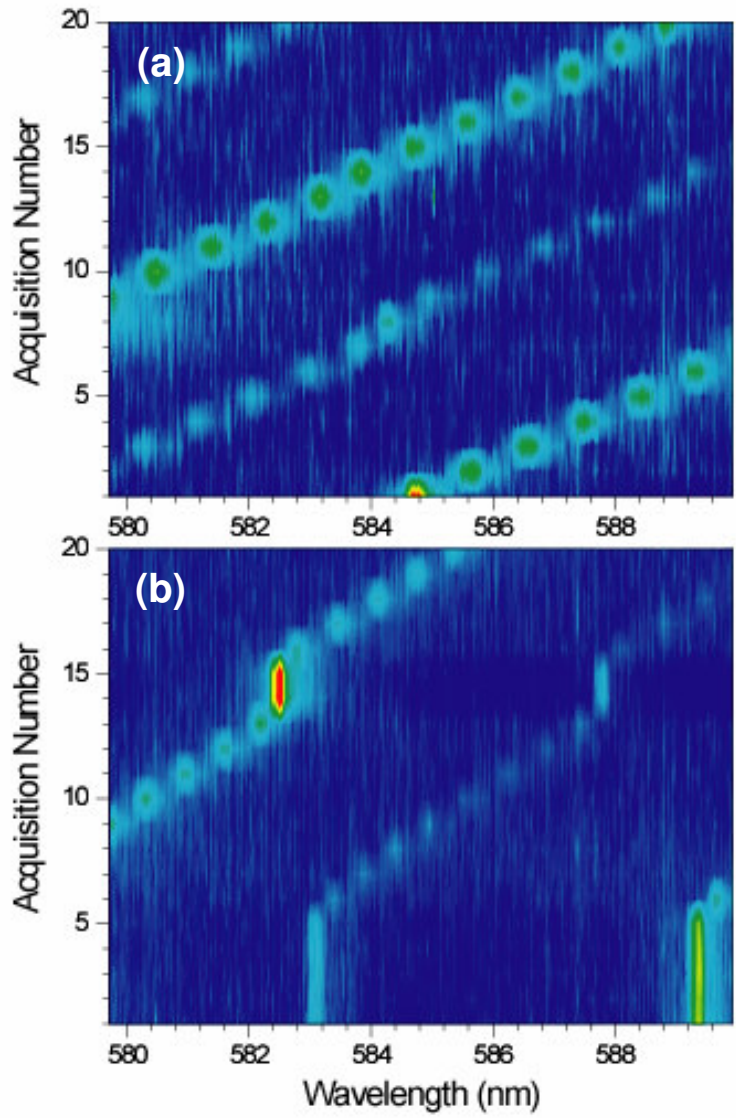